\documentclass[12pt]{article}
\usepackage{amsmath}
\usepackage{hyperref}

\begin{document}

\begin{flushleft}
    {\Large \bfseries Comment on ``The eigenvalues and eigenfunctions of the toroidal dipole operator'' \par}
    \vspace{1em}
    {\large \textbf{G. Vaman} \par}
    \vspace{0.5em}
    {\small Horia Hulubei National Institute for R\&D in Physics and Nuclear Engineering \par}
    \vspace{0.5em}
    {\small \textbf{E-mail:} \href{mailto:getavaman@gmail.com}{\texttt{getavaman@gmail.com}} \par}
    \vspace{0.5em}
    {\small \textbf{Keywords:} toroidal dipole operator, eigenvalues, eigenfunctions \par}
\end{flushleft}

\vspace{1em}

\noindent \textbf{\large Abstract} \par
\vspace{0.5em}
\noindent 
We show here  that the spectrum of the operator $\hat{T}_3^{(\theta)}$ studied by Anghel and Dolineanu (Phys. Scr. {\bf 98} (2023) 015223) is continuous, not discrete, contrary to what is claimed by the authors of the paper to which this comment refers. Consequently, there is no splitting of the energy levels corresponding to this operator that could be observed experimentally.

\vspace{2em}
\noindent
In \cite{anghel}, Anghel and Dolineanu approached the subject of the eigenvalues and eigenfunctions of the toroidal dipole operator for a system consisting of a particle confined in a thin film bent into a torus shape. Although in the abstract the authors announce that they give the analytical expressions for the eigenvalues and eigenfunctions of the $z-$axis projection of the toroidal dipole operator $\hat{T}_3$, the authors study in the body of the paper just the component of this operator along the theta coordinate, which is denoted by $\hat{T}_3^{(\theta)}$ and defined in Eqs. (5) and (6) of \cite{anghel}.

The problem of finding the eigenfunctions of $\hat{T}_3^{(\theta)}$ reduces to solving the differential equation written by the authors in their Eq.(7b), which we transcribe here:

\begin{equation}
    \frac{\mathrm{d}\Phi}{\Phi} = \frac{i \frac{t_3}{C_0} - C_2(\theta, a)}{C_1(\theta, a)} \mathrm{d}\theta,
\end{equation}
where the expressions of $C_0$, $C_1(\theta,a)$ and $C_2(\theta,a)$ are given in Eq. (6) of \cite{anghel}, $t_3$ is the eigenvalue of the operator $\hat{T}_3^{(\theta)}$, $a=R/r>1$ is the ratio between the major and the minor radii of the torus and $\theta \in [0, 2\pi]$.

As the authors of \cite{anghel} noticed, $C_1(\theta,a)$ has two zeros (written by them in Eq. (9c)), which makes the right-hand side of the differential equation to be solved to be singular at two points in the interval $[0, 2\pi]$ denoted by $\theta_0^{(1)}$ and $\theta_0^{(2)}$. For this reason, differential equation (1) must be solved separately over the intervals $[0,\theta_0^{(1)})$, $(\theta_0^{(1)}, \theta_0^{(2)})$ and $(\theta_0^{(2)}, 2\pi]$, the primitives $\mathcal{R}(\theta,a)$ and $\mathcal{I}(\theta,a)$ of $\ln\Phi$ having logarithmic divergences at $\theta_0^{(1)}$ and $\theta_0^{(2)}$.

To integrate Eq.(1), the substitution $\theta \rightarrow s$, $s=\tan(\theta/2)$ is used, which transforms the indefinite integrals to be evaluated into integrals of rational functions. We note that $\pi \in (\theta_0^{(1)}, \theta_0^{(2)})$ for any $a>1$, not only for the particular values of $a$ considered in \cite{anghel}. Indeed, $C_1(0,a)=-2(a+1)^2<0$ and $C_1(\pi,a)=2(a-1)^2>0$, so that $\theta_0^{(1)}\in(0,\pi)$, while the symmetry $C_1(\theta,a)=C_1(2\pi-\theta,a)$ gives $\theta_0^{(2)}=2\pi-\theta_0^{(1)}\in(\pi,2\pi)$. As $\pi \in (\theta_0^{(1)}, \theta_0^{(2)})$ and $\tan(\theta/2)$ is infinite for $\theta=\pi$, it follows that when we integrate Eq. (1) on $(\theta_0^{(1)}, \theta_0^{(2)})$ the indefinite integrals  should be performed separately on the intervals $(\theta_0^{(1)}, \pi)$ and $(\pi, \theta_0^{(2)})$. Each of the indefinite integrals on these two intervals is defined up to an arbitrary additive constant and these constants must be chosen in such a way that the eigenfunction $\Phi$ is continuous in $\pi$. 
In Eq. (8) of \cite{anghel} these additive constants have not yet been chosen in a convenient way to ensure the continuity of $\Phi$; consequently, the plots in Figure 2(b) of \cite{anghel} exhibit a jump in $\pi$, - a jump that the authors denote by  $\Delta \mathcal{I}_{\pi}$ and write in their Eq. (11).

The adjustment of the integration constants ensuring that the eigenfunction $\Phi$ is continuous is performed only in eq. (12) of \cite{anghel}, by adding $\Theta(\theta-\pi)\Delta \mathcal{I}_{\pi}$ to the imaginary part of the eigenfunction. By this operation, the entire portion to the right of $\pi$ in the graphs of Fig. 2(b) is shifted up by the amount $\Delta \mathcal{I}_{\pi}$. In the same equation (12), the authors subtract $\mathcal{I} (0,a)$ and $\mathcal{R} (0,a)$ from the imaginary and real part of the eigenfunction respectively, so that its normalization is convenient, and the so adjusted continuous eigenfunction is denoted by $\mathcal{T}_3^{(t_3)} (\theta, a)$. By subsequently imposing the periodicity condition $\mathcal{T}_3^{(t_3)} (0, a) = \mathcal{T}_3^{(t_3)} (2\pi, a)$, the authors determine the eigenvalues of the operator $\hat{T}_3^{(\theta)}$, eigenvalues which, in their view, are distributed within a discrete spectrum given by Eq. (13) of \cite{anghel}.

It should be noted here, however, that the shifting up of the imaginary part of the eigenfunction over the entire interval $[\pi, 2\pi]$ is not the only possible choice that ensures the continuity of the eigenfunction at $\pi$.  The function (12) of \cite{anghel} is indeed a solution of Eq. (1) on each interval, but it is only a particular one. Since $\theta_0^{(2)}$ is a singular point of Eq. (1), the integration constants on the intervals $(\pi, \theta_0^{(2)})$ and $(\theta_0^{(2)}, 2\pi]$ are independent and need not be related to each other. By shifting the whole interval $[\pi, 2\pi]$, the authors implicitly impose a relation between them, and the periodicity condition applied to this particular choice produces the spurious quantization condition (13) of \cite{anghel}. In fact, to ensure the continuity of the eigenfunction at $\pi$, it is sufficient to shift the imaginary part of the eigenfunction only over the interval $(\pi, \theta_0^{(2)})$, which leads to another solution of Eq.(1), equally continuous at $\pi$:

\begin{equation}
\begin{split}
    \mathcal{T}_3^{(t_3)}(\theta, a) \equiv {} & N(a) \exp \Bigl\{ i t_3 \Bigl[ \mathcal{I}(\theta, a) - \mathcal{I}(0, a) \\
    & + \left( \Theta(\theta - \pi) - \Theta\bigl(\theta - \theta_0^{(2)}\bigr) \right) \Delta \mathcal{I}_{\pi} \Bigr] + \mathcal{R}(\theta, a) - \mathcal{R}(0, a) \Bigr\}
\end{split}
\end{equation}
This eigenfunction of the operator $\hat{T}_3^{(\theta)}$ satisfies the periodicity condition $\mathcal{T}_3^{(t_3)}(0, a)= \mathcal{T}_3^{(t_3)}(2\pi, a)$ for any real eigenvalue $t_3$, which means that the spectrum of the operator $\hat{T}_3^{(\theta)}$ is continuous.

The step functions in (2) generate no spurious terms. The derivative of $\Theta(\theta-\pi)$ gives a term proportional to $\delta(\theta-\pi)$, which exactly cancels the jump of $\mathcal{I}(\theta,a)$ at $\theta=\pi$. On the other hand, $\Theta\bigl(\theta-\theta_0^{(2)}\bigr)$ is constant on each of the intervals $(\pi,\theta_0^{(2)})$ and $(\theta_0^{(2)},2\pi]$, on which Eq.~(1) is solved, so its derivative vanishes there. The only point where it could contribute a delta term is the singular point $\theta_0^{(2)}$ itself, which is excluded from the domain of integration. Moreover, the change of variable $u=\int \mathrm{d}\theta/C_1(\theta,a)$ maps $\hat{T}_3^{(\theta)}$, up to a similarity transformation, onto the momentum operator $-iC_0\,\mathrm{d}/\mathrm{d}u$ on the real line, whose spectrum is continuous.

Consequently, the quantization condition (13) of \cite{anghel} is an artefact of a particular choice of the integration constants, and the spectrum of $\hat{T}_3^{(\theta)}$ is continuous. The eigenfunctions are non-normalizable for every $t_3$ and, as follows from the authors' own Eq. (29), they are naturally normalized to $\delta(t_3-t_3')$ rather than to a Kronecker delta, so that the expansion (31a) of \cite{anghel} should be an integral over $t_3$, not a sum. Therefore, the interaction energy $E_I=\mu_0 e J_z t_3$ of Eq. (14b) of \cite{anghel} varies continuously, there are no discrete energy levels associated with $\hat{T}_3^{(\theta)}$, and the energy splitting $\Delta E_{\rm tor}$ estimated in Eq. (16) of \cite{anghel} does not exist. The experimental signature of the toroidal dipole quantization proposed in Section 2.1 of \cite{anghel} is thus absent.


\begin{thebibliography}{1}

\bibitem{anghel} 
Anghel D-V and Dolineanu M 2023 The eigenvalues and eigenfunctions of the toroidal dipole operator \textit{Phys. Scr.} \textbf{98} 015223

\end{thebibliography}
\end{document}